# Event-Cloud Platform to Support Decision-Making in Emergency Management

Matthieu LAURAS[1,*], Frédérick BENABEN[1], Sébastien TRUPTIL[1], Aurélie CHARLES[1]

[1]Université Toulouse – Mines Albi, Campus Jarlard Route de Teillet, 81000 Albi, France

[2]Université Lumière Lyon 2, Lyon, France

*corresponding author: lauras@mines-albi.fr , +33 (0)5 63 49 32 16, +33 (0)5 63 49 31 81

*Abstract:*
The challenge of this paper is to underline the capability of an Event-Cloud Platform to support efficiently an emergency situation. We chose to focus on a nuclear crisis use case. The proposed approach consists in modeling the business processes of crisis response on the one hand, and in supporting the orchestration and execution of these processes by using an Event-Cloud Platform on the other hand. This paper shows how the use of Event-Cloud techniques can support crisis management stakeholders by automatizing non-value added tasks and by directing decision-makers on what really requires their capabilities of choice. If Event-Cloud technology is a very interesting and topical subject, very few research works have considered this to improve emergency management. This paper tries to fill this gap by considering and applying these technologies on a nuclear crisis use-case.

# 1. Introduction

Imagine a large quantity of radioactive substance is accidentally released in the atmosphere, due to a critical accident in a nuclear plant. To resolve this crisis, a lot of heterogeneous actors may be involved. The services provided by these actors are also diverse and varied, ranging from psychological assistance to traffic duty. This heterogeneity is probably the main cause of the difficulty to manage such an emergency situation. But there are many other difficulties to cope with. (De Maio et al. 2011) have showed that one of the main challenges consists in managing the high amount of data and information from heterogeneous sources (devices, stakeholders, etc.). The Emergency Management (EM) strategy is generally based on complex plans and takes into consideration many factors that may be complementary, contradictory and competitive. It is obvious that crisis situations require tools that can handle all these difficulties to facilitate, in fine, the decision-making.

Our research work consists in developing Information Technology (IT) solutions that could facilitate coordination between actors and support decision-making. (Lee et al. 2012) explain that the key question in disaster context is how to manage for the appropriate information and knowledge quickly and accurately from the massive information and knowledge based on the nature, characteristics, status and situations of the unexpected emergency in order to support the intelligent decision-making process in handling emergencies. (Shaluf and Ahamadun 2006) or (Taohidul Islam and Chik 2011) have demonstrated that advances in IT should be very beneficial for answering these questions. Within our use-case, we have chosen to simulate topical emergency response thanks to a complex-event platform designed and built within the framework of the on-going European research project PLAY (http://www.play-project.eu). The main benefit of using this platform should be that the management of the crisis would be facilitated by the increased situational awareness provided by the platform. In addition to that, the platform would ensure a timely and adequate diffusion of information to relevant actors. All these characteristics make this research work an accurate illustration for a relevant connection between Internet of Things and Internet of Services, and its benefits for emergency management, although it is impossible to imagine this context as a strongly computed environment.

This paper aims at presenting the way Event-Cloud computing could be used to support disaster management. The global objective is to define workflows and web-services simulating crisis management and using the PLAY platform to run and adapt the overall behavior. The paper is split-up in four sections. The first one presents the existing IT systems that are supposed to be relevant to support decision-making in crisis context. The second one presents the proposed platform and its interests. The third one develops the nuclear crisis use-case we propose to test the benefits of such a technology in case of disaster response. The last one proposes a discussion on the forces and weaknesses of our approach for EM.

# 2. Background and Contribution Positioning

## 2.1. General stakes of Emergency Management software

Nowadays, a lot of IT Systems exist and can potentially support crisis/disaster/emergency response processes. These systems are generally called Crisis Information Management Systems (CIMS), Disaster Information Management Systems (DIMS) or Emergency Management Software (EMS). All these systems are used by EM professionals to deal with a wide range of disasters (including natural or human-made hazards) and can take many forms (Lee et al., 2012): (i) training software such as simulators are often used to help for the preparation for the first responders, (ii) word processors can keep form templates handy for printing and (iii) analytical software can be used to perform post hoc examinations of the data captured during an incident. There are particularly used in EM operation centers (crisis cells) to support the management of crisis information and the response processes.

(Lee et al. 2012) have shown that to handle emergencies, crisis stakeholders have to collect the relevant knowledge to advise and/or make timely decisions. The knowledge they want/need are often located in various, numerous and unpredictable sources. In emergency management, decision-makers are confronted with an explosive amount of information that is disseminated among different authorities, external sources (such as press media and web), and other people within a short period of time (Lee et al. 2012).

## 2.2. Crisis Information Management Systems functionalities

In practice the aim of CIMS is to provide a complete suite of IT functions addressing the many requirements from the emergency management community (Iannella et al., 2007). Regarding the research-works of (Iannella et al., 2007; Shankar, 2008; Hiroi et al., 2010; Lee et al., 2012) six groups of tools could be identified according to (see. Table 1):

- (1) The part of the crisis situation concerned by the tools: impacted system; crisis response processes;
- (2) The principal features of the tools: communication; gathering of data; decision support.

| Function of tools / Field concerned | Impacted System | Crisis Response Processes |
|---|---|---|
| **Communication** | 1<br><br>*Examples of Information Systems:* Reliefweb.int, One response, SAHANA, Emergesat, CRISIS$^{TM}$… | 2<br><br>*Examples of Information Systems:* Reliefweb.int,, One response, SAHANA, Emergesat, EM2000, EOC System, OpsCenter, … |
| **Gathering of data** | 3<br><br>*Examples of Information Systems:* CartONG, Parefeu, Responsphere, Rescue, Emergesat, E-Team, LEADERS, … | 4<br><br>*Examples of Information Systems:* Sigmah 1.0, Geophenix-operations, CartONG, E-Team, Country Response Information System… |
| **Decision support** | 5<br><br>*Example of Information Systems:* Country Response Information System, Parefeu... | 6<br><br>*Examples of Information Systems:* **None for the moment.**<br><br>Very few projects are under development such as the SIGMAH 2.0 project (Sarrat and De Geoffroy 2011), PRONTO project (Pottebaum et al. 2011) and of course our own research work: **PLAY project**. |

**Table 1:** Existing Crisis Information Management Systems Categories

A lot of systems able to provide and spread information on the crisis description and/or its management are available (see cells 1 and 2 of Table 1). One of the most representative examples is "Reliefweb" (http://www.reliefweb.int) which is an Internet platform dedicated to communication in a humanitarian disaster context. For each crisis, this platform records raw information in order to inform other stakeholders about the last events. People are consequently kept informed of the crisis situation, its evolution and the ongoing operations. Nowadays, a great majority of crisis decision-centers uses such a system to communicate. Nevertheless, these kinds of systems could be considered as quite basic (with very little added value).

The tools presented in cells 3 and 4 of Table 1 relate to systems that gather data in order to characterize the crisis (victims, damage, position, etc.) or the response operations (means engaged, geographical position, status of operation, etc.). This category of systems is led by a huge number of Geographical Information Systems (GIS). For instance, "CartONG" (http://www.cartong.org/), generating maps on epidemiology, topography, place of crisis during humanitarian crises or "GEOPhoenix-operations" (http://www.geoconcept.com) to follow in real time the position of the emergency vehicles. In this category of systems, a new type of ERP (Entreprise Resources Planning) appears slowly. The Sigmah 1.0 project is one example of this trend (http://www.sigmah.org). This project consists in developing a simple, easy-to-use tool that centralizes and cross-references all the data associated to a crisis response in the humanitarian sector. If Sigmah is limited to the data management functionalities for the moment the consortium wishes to propose a more complete system, which could support the decision-making in crisis situation. A new 2.0 version should be developed in that sense.

The last category of tools relates to the decision support systems in crisis situations (see cells 5 and 6 of Table 1). There are very few tools available in this category today. Nevertheless some systems propose functionalities that simulate the evolution of the behavior of the crisis system. For instance, the "PAREFEU" tool (http://www.isted.com) carries out the simulation of the propagation of a fire. Despite this kind of tool dedicated to the impacted system, no decision-support system dedicated to the response processes seems to exist. However, as discussed in the introduction, there is a real need for tools that can support decision-making in crisis situation, particularly in order to know, rapidly and effectively, which problem needs to be resolved.

We can remark that all these systems are complementary during a crisis situation. And this analysis shows clearly that some functionalities are still missing in existing tools, particularly the decision-support functionalities. (Iannella et al., 2007) confirm this conclusion. Through a CIMS framework composed of 3 layers and 12 functions, the authors list all the functionalities that exist in current CIMS. No decision-support function appears in this framework. Actually, the authors explain that CIMS only propose aggregated reports (text files), budgets, expenditures or geo-spatial images. Due to this lack and given that it constitutes a main need for crisis experts today, we have chosen to focus our research work on this point.

## 2.2. Towards a better emergency management through new IT technologies

Traditional EM response is based on predetermined workflows and disaster plans. This is clearly insufficient to support coordination in emergency response as shown by (Turoff et al. 2004) or (Yu and Cai 2012). The performance of the response is very contingent to knowledge integration, situation awareness and adaptability capabilities (Faraj and Xiao 2006). To address this problem some researchers (few) have recently imagined to use new IT technologies such as Service-Oriented Architecture, Event-Driven Architecture or Cloud-Computing (Huang et al. 2010; Hiroi et al. 2010; Pottebaum et al. 2011; Yu and Cai 2012). Particularly, the use of an event-based approaches offer high potential for coordination support in emergency response operations (Pottebaum et al. 2011).

An "event" is defined as a notable thing that happens inside or outside the studied system, as for instance: a problem, an opportunity, a threshold, a deviation, etc. The most important issue is an efficient detection of such events, leading to the co called situational awareness. As evidence, this notion fits clearly with emergency situations and "event-driven" approaches should be very useful to improve crisis response and management. (Yu and Cai 2012) indicate that the existing event-based systems commonly employ a publish/subscribe interaction paradigm, where actors have the ability to express their interests in a set of events or patterns of events, in order to be notified subsequently when any event that matches their subscribed interests is published. Nevertheless the authors affirm also that it becomes impractical for EM users to pre-determine the relevance of all possible events and reason about impacts on their activities. Then to scale up effectively the event-based approach to emergency situations, (Yu and Cai 2012) suggest integrating the event-driven architecture with complex event-processing (CEP) engines to provide more effective event management in response operations.

Nowadays two main research works have tried to propose such a mechanism:

- The PRONTO project has proposed some methods and results for the identification, definition and validation of events that happen in EM situation and corresponding to the event objects that could be processed by information systems (Pottebaum et al. 2011). If this contribution is particularly useful, it does not constitute a concrete operational tool for EM practitioners (this is only a framework).

- The (Yu and Cai 2012) research project addresses the problem of cognitive aid to human actors in crisis situation through an event-driven approach. Their proposal contains two main contributions: (i) it represents and maintains a group mental model of the overall collaborative activities and their interdependencies; (ii) it helps the users to determine the relevance of events by automating the reasoning on how an event impacts the states of activities as its effects propagate through the web of dependencies. Although this proposal was very interesting, some limitations appear. Particularly, the proposed CEP is connected directly to the different event providers (devices). This approach requires structured data and limits the interoperability capabilities regarding the heterogeneity and the distribution of real systems (IS and devices). Another limitation is about the incapability of this system to ensure the control of the EM business processes. Actually, the proposition focused on situational management (awareness), not on business process management. It should be interesting to propose complementary functionalities able to support the orchestration[1] and the choreography[2] between the stakeholders of an EM response. Finally, the proposition of (Yu and Cai 2012) does not really develop solutions for disaster practitioners to respond more quickly and more adequately to short-term problems, disruptions and changes.

Based on this, our problem statement consists in designing a new IT platform in line with the philosophy of event-driven approach and with the results of (Pottebaum et al. 2011) and (Yu and Cai 2012) but that addresses their main weaknesses. This proposition is developed in the following section and contains three main contributions regarding previous research-works:

- Ability to connect to heterogeneous and distributed systems through Cloud Computing technologies.

- Ability to orchestrate and to choreograph EM business processes through Distributed-Service Bus.

- Ability to detect and suggest adaptations of the response regarding current situation through Service-Adaptation Recommender.

## 3. Proposed Emergency Management Event-Cloud Platform

### 3.1. Main functionalities of the proposed platform

The PLAY platform is a Web-oriented structure to combine events from many sources with the goal of connecting and orchestrating services, devices and people as shown in Figure 1.

---

[1] Orchestration is the ability to manage the business processes of an organization.
[2] Choreography is the ability to manage the collaborative business processes between two or more organizations.

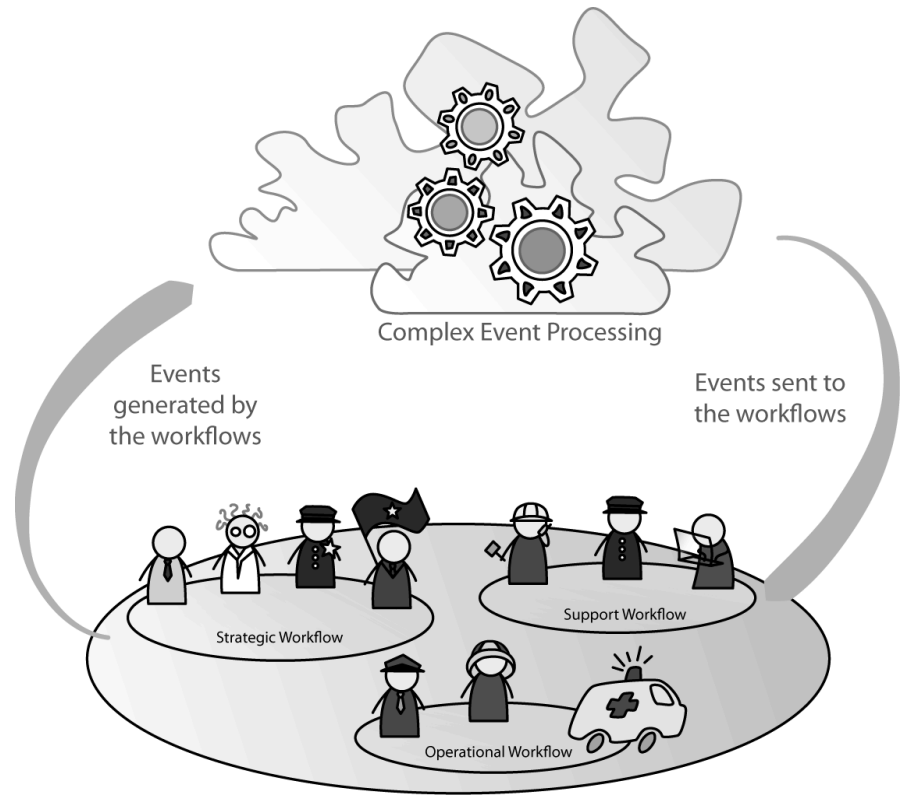

**Fig. 1:** Nuclear Crisis Use Case, Actors and Events

The platform has emerged as an event marketplace, a place that brings together the senders and receivers of events and provides numerous services on top of them. To that end PLAY combines several technologies to deal with delivery, processing and storage of events as real-time information. We will briefly outline these technologies by introducing the components of the platform in Figure 2. The reader could access to the detail of the conceptual architecture of the PLAY platform on the dedicated website (http://play-project.eu).

The Distributed Service Bus (DSB) provides the Service-oriented Architecture (SOA) and Event-driven Architecture (EDA) infrastructure to connect components, devices and end user services (through the Information Systems of the stakeholders). The DSB aims to provide connectivity between services providers, services consumers, event consumers and event providers, potentially distributed over distinct administrative domains, in a completely transparent way for the user point of view. Thus, distributed sources of events can be combined in the platform.

The Event Cloud provides storage and forwarding of events so that interested parties can be notified of events according to content-based subscription. The storage operates as an event history to fulfill queries for older events, which do not need real-time results e.g., when generating statistics. The Event Cloud is comprised of a peer-to-peer system of storage nodes organized in a controller area network (Filali et al. 2011).

The Distributed Complex Event Processing (DCEP) component has the role of detecting complex events and does reasoning over events in real-time. The main issue in the event-driven computation is to provide the right information to the right people/components in the right situation. Events might not be meaningful, but meaningful events can be derived from available, simpler events. The platform can readily detect such derived events, because it has knowledge of all events and applies event patterns, as described in (Etzion and Niblett, 2010), to the input events.

Finally, the Service Adaptation Recommender (SAR) suggests changes (adaptations) of services' configurations, composition or workflows, in order to overcome problems or achieve higher performance. The objective of the Adaptation Mechanism is supporting the decision by suggesting changes (adaptations) of services' configurations, composition or workflows. Based on recognized situations, this mechanism will be able to define and detect adaptation "opportunities", proposing adaptation actions to the end users of service based applications involved in workflows. At the same time, it will undertake the responsibility of revealing to them the reasoning process that led to the adaptation recommendation, providing them with the capability of accepting or rejecting the proposed alterations.

Events in PLAY may originate from diverse devices, services and users such that a versatile event format and matching query language are required. To deal with this heterogeneity we propose an event format based on Resource Description Framework (RDF) with a matching SPARQL-based event pattern language syntax. Both base-technologies RDF (Klyne and Caroll, 2004) and SPARQL (Harris and Seaborne, 2010) are currently used on the Web as general methods for conceptual modelling (and querying, respectively) of information. We are adapting them to enable a real-time Web based on well-known foundations i.e., RDF and SPARQL.

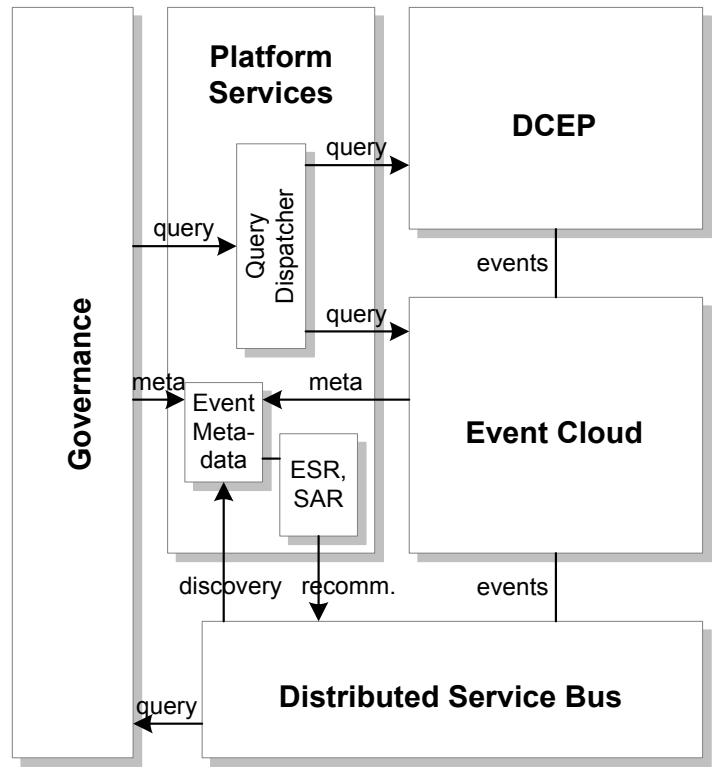

**Fig. 2:** Conceptual Architecture

Therefore the Complex Event Processing has all the information of the crisis response and evolution can be detected thanks to rules. Once an evolution is detected, an alert event is sent to the right stakeholder at the right time. When this kind of event is received, a decision of modification could rapidly be made and executed.

## 3.2. Usefulness of the proposed platform for emergency management

In crisis circumstances, the proposed Event-Cloud platform would be considered two complementary levels: (i) on the decision level, linking actors of that layer in order to help them to adjust the crisis reduction processes according to identified events, and (ii) on the field level, dealing with orchestration and choreography of services. All the expected partners on the field (such as firemen, police, army…) should provide events (e.g. reports concerning their actions as well as measures concerning the observed situation).

Such a situation provides some representative characteristics regarding the three contributions we want to address in this research-work (see. Section 2):
- A lot of heterogeneous actors, possibly widely distributed, may be involved (and so a lot of associated services) and this heterogeneity is one cause of the difficulty to manage crisis situations;
- A lot of critical dependencies between the actions of these heterogeneous actors (collaborative processes describe the chronology of activities but also how activities might pre-condition or post-condition for each others);

- Crisis situations are obviously the kind of context where agility (especially responsiveness and flexibility) is one critical point. It is crucial that workflows and actions remain perfectly adapted to the possibly changing situation.

All these characteristics make the use of Event-Cloud technologies a relevant way for improving EM. Actually, the multiplicity and diversity of actors involved, the volume and heterogeneity of information, the critical dependencies between actions as well as the dynamics of the situation make the situation very complex. The challenge consists in developing a solution able to support stakeholders to drive such a complex situation and decision-makers to be more agile. The main objective should be relieving decision-makers of inaccurate or irrelevant information.

In order to demonstrate the relevance of our system regarding EM, we have decided to define a representative scenario (See. Section 4), to test it on a specific nuclear crisis scenario (see. Section 5), and to discuss the advantages and limits of our system (See. Section 6).

## 4. Use-Case Presentation

### 4.1. Gathering, structuring and modeling the knowledge

The proposed use-case is based on French legislation recommendations and predefined emergency plans (*Plan d'Urgence Interne* and *Plan Particulier d'Intervention*) regarding nuclear crisis situations. French Nuclear crisis experts (firemen, policemen, representative of national authority, etc.) have amended the use-case in order to maintain the realism and the relevance of the scenario.

Concretely, we are considering a situation in which nuclear plant teams detect a leak between primary and secondary loops, thanks to the alert given by a high-pressure sensor in the primary loop. The throttle valve is open and does not respond to closing order, so the teams realize that there is a risk of radioactive leakage in the atmosphere (see. figure 3). The responsible of the nuclear plant then informs the representative of the national authority, who activates the Emergency Plan in reflex mode. A crisis cell is settled under the responsibility of the *Representative of National Authority*. The crisis cell alerts field actors (firemen, police, army, office of infrastructure…) and ask the radiation survey network (RSN) and weather forecast institution (MF) for measurements. It alerts the media and set off the siren so that the population can learn that they have to stay indoors and listen to media. Field actors are deployed to execute dedicated operations to control the situation and limit the negative impacts of the crisis.

Based on the knowledge included in the official emergency plans, nineteen business processes have been identified, modeled and finally validated by experts of nuclear crisis. For better structuration of the knowledge, these business processes have been dispatched into three complementary levels accordingly to ISO 9001:2008 standards (see. Figure 3).
1. Strategic level - To manage nuclear crisis: This first level is dedicated to present the decisional part of the process cartography. It concerns decision making during the crisis management. Practically, 6 decisional processes (1 in opening phase, 4 in regular phase and 1 in closure phase) were identified. They are dedicated to take decision and deal with overall management, such as "protect population" or "manage situation".
2. Operational level - To resolve nuclear accident and its consequences: This second level deals with the concrete operational part of the process cartography. It concerns mainly the actions performed on the crisis site. 9 operational processes (1 in opening phase, 7 in regular phase and 1 in closure phase) were also determined. They are dedicated to perform the concrete field activities that should be done to reduce the crisis situation, such as "confine population" or "implement circulation plan".
3. Support level - To support nuclear crisis response: This last level concerns the supporting activities dedicated to provide means to other processes and to ensure logistic aspects of the crisis management. Finally, 3 support processes (all along crisis response) were underlined. They are dedicated to provide others processes with necessary data and resources, such as "assess situation" or "manage resources".

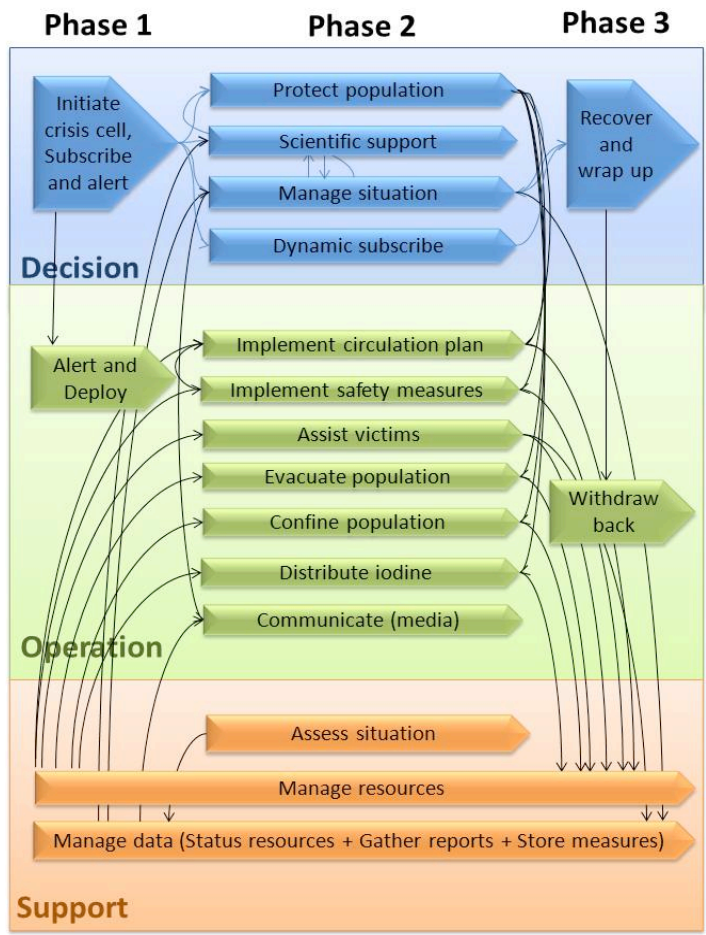

**Fig. 3:** Business processes cartography

All these processes have been described in detail through a *Business Process Modelling Notation* language as suggested by (Juric et al. 2006) (see. http://www.bpmi.org to access to this international standard). As examples, the following figures present two business process descriptions: manage situation and assess situation. On these diagrams, pools (global containers) or swim lanes (sub-containers in one pool) represent the involved actors: representative of national authority, weather forecast institution (MF for "Météo France"), radiation survey network (RSN) and the PLAY system. Each pool embeds its own activities and flows, while exchanges between pools are represented through flows generating events.

The first business process (Figure 4) fits with a process involving representative of national authority in charge of making decisions in order to limit the consequences of the crisis. This process starts when the crisis-cell is settled. Then the representative of national authority can consult available data or wait for new events regarding weather (Alert MF), radioactivity (Alert RSN) or field situation (Alert field). Based on this information, the decision-maker will analyze the situation and select one option among: (i) close the crisis operation, (ii) ask for activity report from the field operations, study the report, and then make a decision, (iii) ask for an advice to nuclear specialists, study the advice, and then make a decision, or (iv) make directly a decision. This business process is executing continuously while the crisis cell is activated.

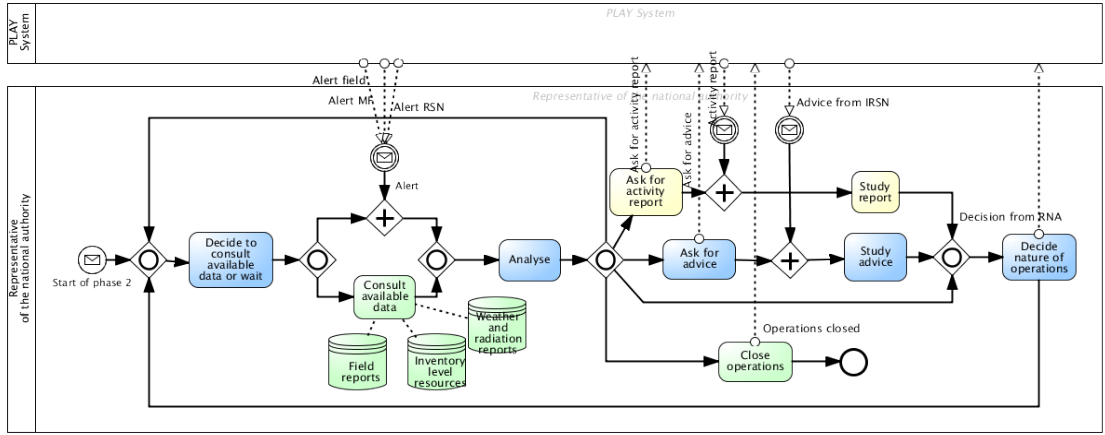

**Fig. 4:** Situation Management Workflow

The second business process (Figure 5) fits with a support process in charge of delivering radioactive and weather measurements on a continuous way. Those processes start when the crisis occurs. The principle of these activities consist in measuring the radioactivity or the weather conditions (through sensors on the field) and to send the measurements to the PLAY platform every 30 seconds. All these data might be gathered, analyzed and distributed to subscribers by the platform. Regarding the radioactivity measurement, the representative of national authority can decide to extend the perimeter under surveillance by activating new sensors. These business processes stop when the crisis operation is closed.

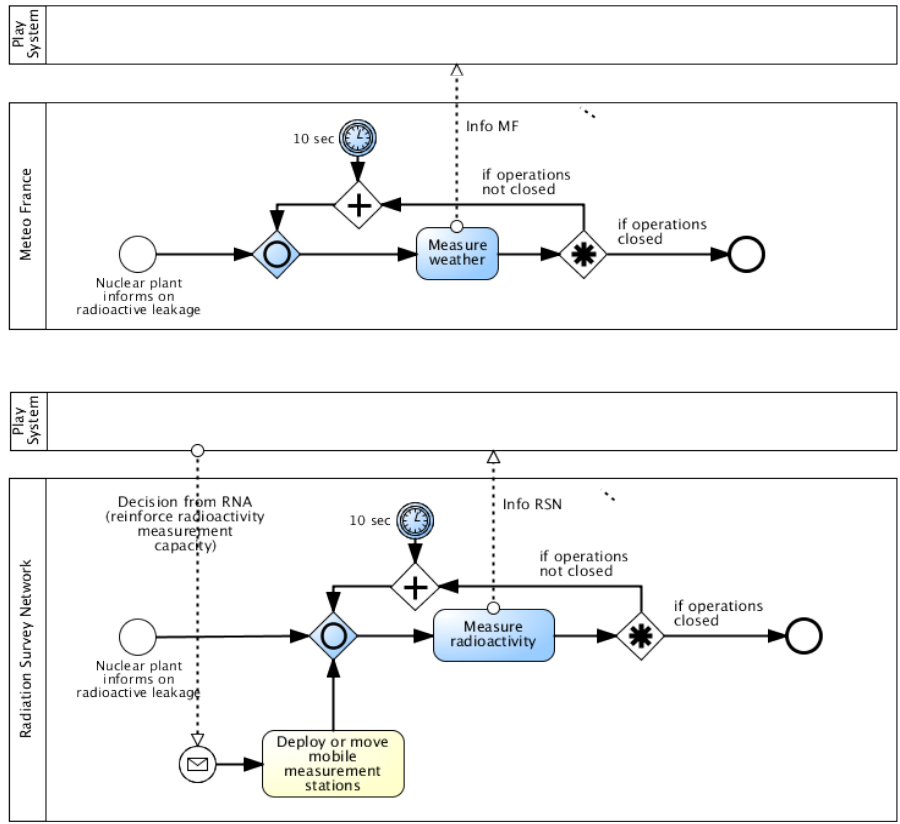

**Fig. 5:** Assess Situation Workflow

## 4.2. Establishing a representative and illustrative EM scenario

Even if more complex and more complete scenarios have been set up and simulated during the European PLAY project (see http://www.play-project.eu for more information on them), we have chosen to develop in this paper only a representative one. This scenario is especially based on the workflows described on Figure 4 and Figure 5.

The challenge of this scenario is to underline the capability of PLAY platform to support efficiently an EM situation.

Regarding, the following Figure, this scenario is based on 6 sets of business processes on the 18 that are included in the whole scenario (see. Figure 3). The Figure 6 shows an overview of the scenario.

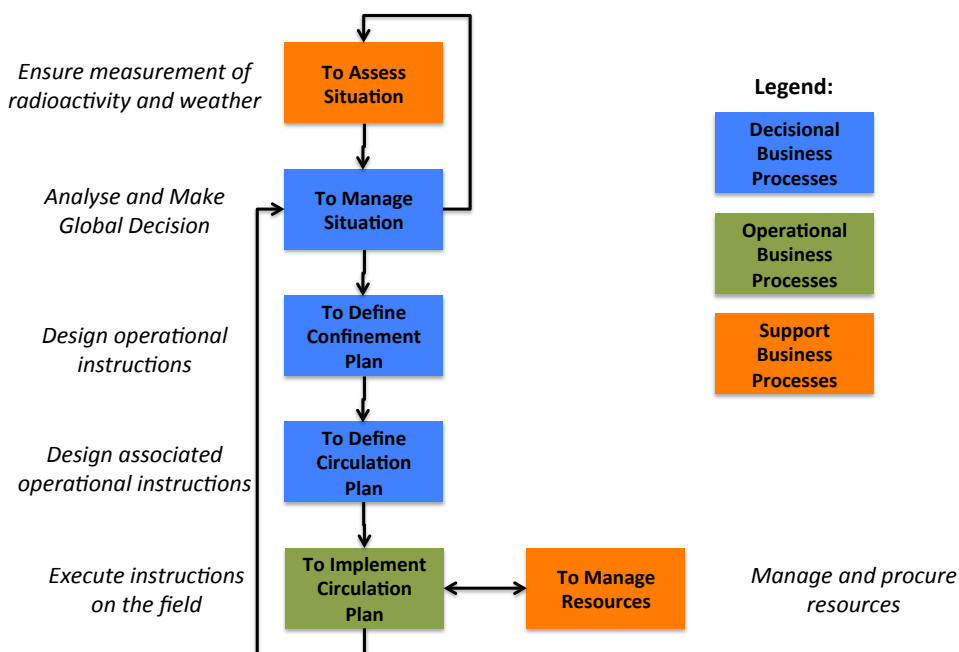

**Fig. 6:** Scenario overview

At the beginning, the crisis-cell is set-up following an accident (radioactivity substance in the atmosphere) detected by the nuclear plant operator. In this scenario, we will consider the following chronological events:

- Storyboard of the 1$^{st}$ period: The representative of National Authority (Préfet) will use the PLAY platform to support analysing of the huge amount of radioactivity and weather measures available. Based on this, he/she will be able to make some decisions regarding the perimeter that has to be under-surveillance. He/she will be able to ask some advices though the platform to French Nuclear Expert Institution (IRSN) to support his/her decisions (see. Figure 4).
  *Objectives of the 1$^{st}$ period: This first part of the scenario will be used to assess the scalability of the PLAY Platform and the ability to filter and produce added-value information for decision-makers (through CEP treatments). This part of the scenario will demonstrate how the platform could facilitate the collaboration between stakeholders.*

- Storyboard of the 2$^{nd}$ period: Based on different measurements gathered, the PLAY platform will alert the *Préfet* that a confinement should be engaged. After a brief analysis of the situation the *Préfet* will decide to activate the confinement plan. The decision will be sent to other crisis stakeholders through the PLAY system. Then, the representative of the Police will design the confinement plan and transmit it to other stakeholders through the Platform. To be executed a confinement plan supposes different things such as (i) informing population to stay at home, (ii) distributing iodine capsules to people who potentially does not have, (iii) securing the area to avoid panics or malevolence acts, and

(iv) preventing entrances of new people in the concerned area. In this sub-scenario, only the last business process is managed. The representative of Office Infrastructure will design its own operational plan (circulation plan) based on the confinement plan established previously by the representative of the Police. When the circulation plan is ready, it is sent through the platform to the office infrastructure teams in order to be applied on the field.
*Objectives of the 2$^{nd}$ period: This second part of the scenario will be used to assess the capability of the PLAY Platform to manage coordination (if task 1 is activated then task 2 should/must be activated) and propagation of structured data such as confinement plan here.*

- Storyboard of the 3$^{rd}$ period: During this last part of the scenario, the circulation plan implementation will be done on the field and tracked by the platform. But as for any real situations, different hazards will occur and disturb the process execution. Firstly, some required resources will have been proved insufficient to execute the plan. This problem should be detected by the Platform and the Adaptability component should propose an alternative to follow the process. Secondly, the PLAY system will detect that a task is too long regarding the standard and will activate an alert to decision-makers in order to adapt the decision. Thirdly, the management of resources (trucks, materials, people, etc.) will suppose to manage inventory by checking all resources movements. This point will permit to the office of infrastructure to adapt the circulation plan implementation to the real state of the resources.
*Objectives of the 3$^{rd}$ period: This third part of the scenario will be used to assess the capability of the PLAY Platform to track the execution of different tasks and consequently to detect potential problems. Moreover, this part of the scenario will allow testing the adaptation capabilities of the Platform. Finally, the Platform capabilities in terms of data storage and exploitation will be validate here.*

# 5. Test-Run

## 5.1. 1st period

Our use case is based on French legislation recommendations that impose particularly the following barriers. Note that mSv (millisievert) per hour is the unit of the international system to measure the impact of radiation on human-beings:
- To control the zone if the dose rate is above 0,025 mSv/h.
- To confine and distribute iodine capsules if the dose rate is above 2 mSv/h.
- To evacuate if the cumulative dose rate is above 50 mSv/h.
- To ingest iodine capsules if the cumulative dose rate is above 50 mSv/h.

In this first period of the test, we have considered that the nuclear crisis had already started and that all relief stakeholders had already involved in the crisis cell and on the field. In the crisis cell, the representative of national authority (*Préfet*) managed the situation according to the *situation management workflow* (see. Figure 4). At the beginning of the scenario (defined as timestamp t0), a perimeter of 5 km around the nuclear plant was under surveillance. Into this perimeter, 5 radioactivity sensors were sending a radioactivity measure every 30 seconds, corresponding to 10 events / minute (according to the workflow describes on Figure 5). These events included information about the identity of the sensor, the geographical position of the sensor, the date of the measure and the value of the measure. At a same time, *Météo France* (French weather forecast institution) sent information about the wind situation (see. Figure 5). These measures were based on 5 sensors on the perimeter of 5 km around the nuclear plant. To each weather measure, 2 events were associated. The first one included information about the identity of the sensor, the geographical position of the sensor, the date of the measure and the speed of the wind. The second one included information about the identity of the sensor, the geographical position of the sensor, the date of the measure and the direction of the wind. Each sensor sent these information every 30 seconds, corresponding to a total of 20 events / minute. During the first 3 minutes of the scenario, all values were right regarding the business rules (radioactivity < 1 mSv and no wind). After 3 minutes, radioactivity measures increase progressively (trend = 0,3) up to 1,8 mSv at t0+7 minutes. Regarding the following business rule, the PLAY platform has sent an alert to the *Préfet* (as shown on Figure 7):

> IF *radiation measure* > *V+* OR (*radiation measure* > *V-* AND *dRM/dt* > *s*) → *Alert-RSN*
> If radiation measure exceeds V+ or if radiation measure exceed V- and increase to strongly, then
> send an alert concerning radiation survey network (nota: V+>V-)
> V+ = 2 mSv, V- = 1 mSv, s=0,2

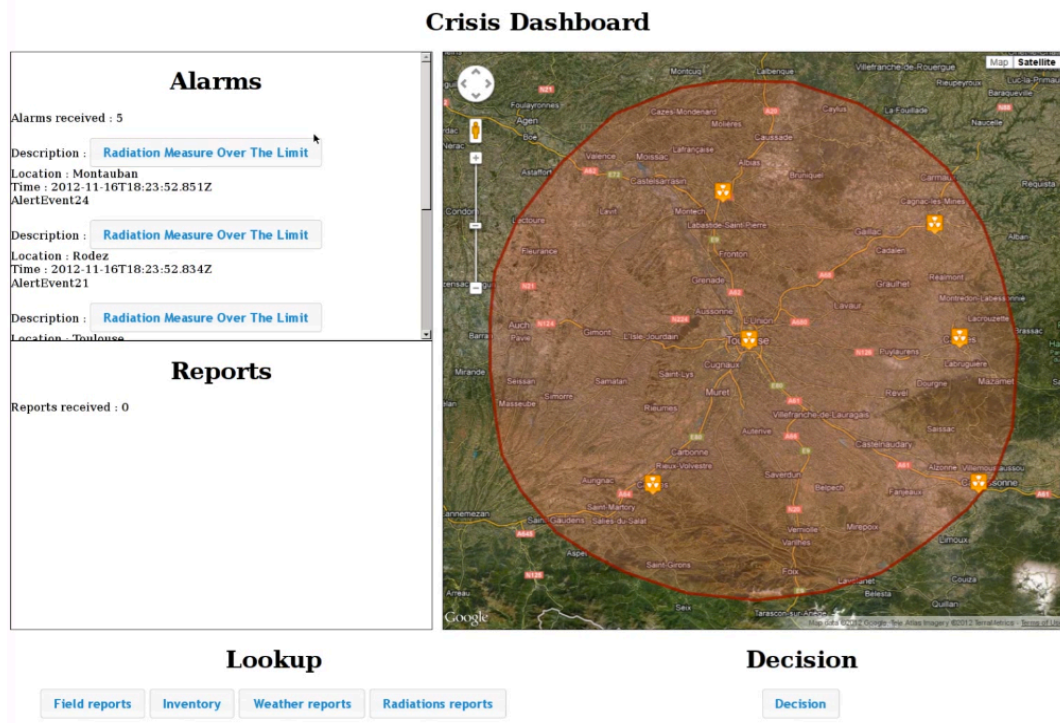

**Fig. 7:** Screenshot of the test-run dashboard

Values of different sensors stayed globally stable between t0+7 minutes and t0+9 minutes. At t0+9 minutes, weather information indicated that the wind rises (40 km/h) in a particular direction (southeast). Regarding the following business rule, the PLAY platform has sent a new alert to the *Préfet*:

> IF *dWindDirection/dt* > *dWD* OR *dWindIntensity/dt* > *dWI measure* è *Alert-MF*
> If the wind change too drastically in intensity or in direction,
> then send an alert concerning meteo france
> dWI measure = 30 km/h

Based on this alert, the *Préfet* decided to extend the perimeter under surveillance up to 30 km in southeast direction. Consequently, 20 additional radioactivity sensors were activated. At this time, the number of events reached 70 events / minute. At a same time (i.e. t0+9 minutes), the *Préfet* asked for advice to IRSN (French Institute for Radio-protection and Nuclear Safety). 5 minutes later (i.e. t0+14 minutes), the report (text) of IRSN was sent by the platform. Based on this information, the *Préfet* decided to extend the surveillance to the whole regional territory in terms of radioactivity surveillance. Then, complementary to the 5 weather sensors activated, a total of 320 radioactivity sensors were ongoing. The total number of events attained 660 events / minute. The first period of the test stopped at t0+20 minutes. Moreover, all along the test-run, the following business rule had been managed by the platform in order to support decision-making.

> Every *5* minutes → Draw graph with last *5* minutes *RSN* measures + *RSN measure graph*
> Every 5 minutes, draw the graph describing last 5 minutes RSN measures and send it as an event

### 5.2. 2nd period

During the 5 last minutes, 3 sensors have sent radioactivity measure higher than V+. The first business rule has been applied again. Consequently, the *Préfet* decided at t0+20 to confine the population on a perimeter of 5 km around the nuclear plant. The following business rule was applied through the PLAY platform according to the French nuclear crisis plans.

> IF *Confinement_Decision* → *Alert-Police_Representative*
> If a confinement decision is made then send an alert to the representative of Police in order to define the confinement plan.

Between t0+20 and t0+25 minutes, the Police representative defined the *confinement plan.* Then, this plan was transmitted to the different subscribers. This plan included information on the precise limits of the confinement and listed the different actions that have to be done. Among them, the office of infrastructure had to organize a security perimeter by closing some roads and by establishing deviations. Accordingly to the French nuclear crisis plans, the following rule has been activated by the PLAY system.

> IF *Confinement_Plan_Validated* → *Alert-Office_of_Infrastructure_Representative*
> When the confinement plan is validated then it has to be transmitted to the representative of office of infrastructure. He/she will have to develop a dedicated circulation plan to avoid new entrances in the confined perimeter.

As discussed in the concerned emergency plans, the *confinement plan* should also activate other decisions such as *to secure perimeter* (Army and Police)*, to distribute complementary iodine capsules* (Firemen and Municipalities)… But the test-run only focused on representative of office infrastructure decisions. At t0+30 minutes, he/she designed the operational *circulation plan* that had to be implemented on the field. This plan indicated that 8 roads had to be closed and 12 deviation tours had to be settled on the perimeter of 5 km around the nuclear plant. This *circulation* plan was sent to the different subscribers in the crisis cell (*Préfet*, Police, Army and Firemen representatives) and on the field (office of infrastructure teams). This second part of the test stopped at t0+30.

## 5.3. 3rd period

This last period of the test is defined between t0+30 minutes and t0+105 minutes. The focus is done on two sets of workflows (not presented in this paper but available on the play project website http://play-project.eu), which are *to implement circulation plan* and *to manage resources*, just regarding the office of infrastructure's resources).

During this period, the office of infrastructure field team has consulted the inventory level of available resources through the PLAY platform and has analyzed the circulation plan sent by the representative through the platform also. These tasks were executed between t0+30 minutes and t0+35 minutes. Based on this information (4 vehicles are available and three different geographical zones have been defined in the circulation plan), the team asked at t0+35 minutes for 3 vehicles armed with generic set of road signs to implement the circulation plan. At t0+35, the person in charge of inventory management at office of infrastructure has received the request through the PLAY system. The demand was validated and availability for t0+40 minutes was confirmed. The implementation started at t0+40 minutes with the three vehicles delivered and was programmed to finish at t0+70 minutes. But at t0+52 minutes, an alert field was sent to the PLAY platform because one of the vehicles had burst. Regarding the following rule, the adaptability component has checked every 10 minutes the potential gaps that could exist between the theoretical model (3 vehicles are involved on the field) and the situational model (2 vehicles were concretely involved on the field).

> IF *Committed_Ressources* <> *Requested_Resources* → Propose different alternatives
> *Ask_For_A_New_Resource* OR *Dispatch_Residual_Tasks_On_Remaining_Resource*
> If there is a difference between the needs and the involved resources, then send an alert to the person in charge of the workflow to inform on problem and propose alternative solutions.

Consequently, at t0+60 minutes, the system detected the problem and proposed to the office of infrastructure team *to request a new vehicle* or *to dispatch the remaining activity on remaining vehicles*. The second option was validated and the information was sent to the two vehicles that stay on the field.

At t0+80 minutes, the implementation of the circulation plan was ongoing yet. No information has been filled from the field teams. Consequently, at this time for the crisis cell members, the implementation of the circulation plan was done (theoretically finished at t0+70 minutes). But the adaptability component checks every 10 minutes the evolution of the running activities. As a reminder all activities connected to the platform are tracked through three potential statuses: (i) waiting, (ii) ongoing, (iii) finished. Consequently, at t0 + 80 minutes, the following rule was applied.

> IF *Current_Activity_Status* <> *Intended_Actvity_Status* → Propose different alternatives *Require_For_Immediate_Reporting* OR *Send_Someone_On_The_Field* OR *Wait*
> If there is a difference between the current status of an activity and the intended status for this activity, then send an alert to the person in charge of the workflow to inform on problem and propose alternative solutions.

Based on these propositions, the representative of office of infrastructure decided to require for an immediate reporting. The field teams sent a short report at t0+83 minutes to explain that they will finish soon. At t0+88 minutes, the implementation of the circulation plan was completed and a final report was sent to the subscribers. The vehicles were released at t0+105 minutes and the inventories were updated.

## 6. Discussions

Through the instantiation of the scenario, we have shown how our proposition can contribute to the management of information and decision in crisis management context. As (Lee et al. 2012) explain, in EM, decision-makers are confronted with an explosive amount of information and they have to apply technological systems to manage them more effectively. The main benefits of the use of our Event-Cloud technologies to support management of crisis are:

- *Eliminating superfluous, inaccurate or irrelevant information*: The use of the Event-Cloud platform has allowed minimizing drastically the number of information circulating between stakeholders. For instance, during the first 7 minutes of the scenario, in a classical configuration, the representative of national authority should receive more than 200 raw measures of radiation and weather. With our proposal the decision-maker can subscribe only to useful information for him/her. In our case, he/she has just received 1 report event every 5 minutes and 1 alert event when necessary (in that case at t0 + 7 minutes). This is not only a classical information aggregation environment (able to provide synthesis instead of raw measures) but also a smart system able to send message specifically on purpose.
- *Automating some analysis or actions based on predefined business-rules:* In addition to the previous filtering contribution that allows relieving the information flow, the unique events received by the decision-maker constitute a real added value for him/her. The report events for instance give a synthetic view of the past situation (already constructed) and should support complementary human analysis and interpretation of the situation. Executing predefined business-rules produces the alert-events. These events give alert on potential risks and allow anticipating problematic situations (alert-event on potential excessive radiation due to the trend of past radiation measures for instance).
- *Reducing the time of information transmission between devices, stakeholders and decision-makers:* The use of proposed technologies allows connecting directly all devices and actors involving on the field and inside the crisis cell. Then, information (events in our case) circulates quasi instantaneously between the actors. During our tests, within a very distributed environment (more than hundred of kilometers between the systems), each event was distributed to all targeted actors in less than 1 second (between 700 and 900 milliseconds). Even if this feature strongly remains dependent on the ability to maintain electronic networks operating, this is nevertheless a nice way to integrate all devices (including people personal devices) in crisis management.
- *Increasing the reliability of information:* The use of technologies such as Enterprise Service Bus allows connecting heterogeneous information systems and devices will very low integration constraints. Within these technologies, no additional treatment or manipulation is needed to transmit or use. The information stays totally intact. Risks of non-quality are consequently drastically reduced. This contribution has been used during the $2^{nd}$ period of the test-run.

- *Improving the agility capabilities of the crisis stakeholders:* The proposed system allows developing the ability of disaster practitioners to respond more quickly and more adequately to short-term problems, disruptions and changes. According to (Charles et al. 2010) this ability can be defined as "agility". Particularly, our proposition brings solution to improve visibility, velocity and reactivity during the execution of the response as demonstrated during the 3$^{rd}$ period of the test-run.

These points fit perfectly with the analysis of (Ibrahim and Allen 2012) who argue that better information sharing plays a crucial role in instilling or enhancing trust and that in the time-bound, uncertain, and highly volatile context of emergency response, if trust collapses, then it must be rebuilt swiftly and this can be done through more effective information sharing. (Bharosa et al. 2010) have shown that disaster practitioners are often more concerned with receiving information from others than with providing information to others who may benefit. Our proposal contributes to limit the weaknesses associated to this fact. (Preece et al. 2013) explain on the other hand that there is a gap in methods for analyzing information and making decisions whilst delivering rapid response. Regarding this analysis, we can affirm that the proposed Event-Cloud techniques should contribute to develop concrete solution to bridge this gap by facilitating the coordination between stakeholders by connecting them efficiently.

## 7. Conclusion and future works

Although it is not necessary a strongly computed environment, several elements make this research work a relevant illustration of the usefulness of Internet of Services for EM situations:
- The heterogeneity of actors and Information Systems or devices;
- The high volume of heterogeneous information;
- The continuous changes in orchestration (internal business processes) and in choreography (interactions between processes) of crisis response processes;
- And the high-pressure environment that imposes to focus capabilities on complex decisions that require human intelligence.

Practically, the proposed platform provides an event management environment, which allows users to be dynamically connected to the existing information systems, devices and other sensors. It is an overall structure of information sharing that selects the right information, for the right person at the right time. Proposed mechanisms allow the Event-Cloud platform to filter, aggregate, and deduce events from received ones. Concerning this point, one have to admit that "being well informed" is nowadays one crucial issue for EM. This is mainly because more and more information exists and is available in crisis context (which was not the case some decades ago) but it is quite difficult to find it and to use it (especially because of the very huge amount of available information). One can simply thinks about emails and how difficult it is to deal with the number of email received per day. This event management, and particularly business rules that may be used to deal with these events in order to produce the right information, can be seen as a very promising opportunity to increase the agility capabilities of EM decision-makers. Finally, such a platform is a solution to deal with agility but also with complex processes management by ensuring responsive coordination tasks (including adaptation tasks). By this way, human beings (and there decision skills) are free from tasks that do not require their human being skills, and could focus on crucial decision tasks.

Though the PLAY platform constitutes a significant first step towards supporting coordination and decision-making in EM situation, the proposition is for the moment only based on a simulated environment of the crisis. To cope with this limitation, we will in our further research carry out experiments over real and live use cases. This will enable us to validate the relevance, the significance and the performance of our system in a provable context. Actually, our proposition needs to be studied regarding its sensitivity to the heterogeneity of systems and the variability of response business processes that involve during the crisis. Further research should also include complementary studies regarding the prerequisites for implementing efficiently such a solution as: (i) capability to model (even in a raw manner) the EM business processes; (ii) availability of the telecommunication network during a crisis; (iii) capability to gather and share "events" (activity status, sensor measures, reports, etc.). It will also be interesting to assess the cultural impact of such a system on EM stakeholders in order to observe probable cultural oppositions. All these perspectives are geared towards real applications (exercises, trainings and real situations) within the French Interior Ministry.

# 7. Acknowledgement

The PLAY project (Pushing dynamic and ubiquitous interaction between services Leveraged in the Future Internet by ApplYing complex event processing) is being funded by the European Commission under Seventh Framework Program (Grant FP7-258659). The authors would like to thank the project partners for their advices and comments regarding this work.